# Performance Benefit of Aerocapture for the Design Reference Mission Set

Athul Pradeepkumar Girija [1***]

[1]*School of Aeronautics and Astronautics, Purdue University, West Lafayette, IN 47907, USA*

**ABSTRACT**

Aerocapture is a maneuver which uses aerodynamic drag to slow down a spacecraft in a single pass through the atmosphere. All planetary orbiters to date have used propulsive orbit insertion. Aerocapture is a promising alternative, especially for small satellite missions and missions to the ice giants. The large ΔV requirement makes it practically impossible for small satellites to enter low circular orbits. Aerocapture can enable insertion of low-cost satellites into circular orbits around Mars and Venus. For ice giant missions, aerocapture can enable orbit insertion from fast arrival trajectories which are impractical with propulsive insertion. By utilizing the atmospheric drag to impart the ΔV, aerocapture can offer significant propellant mass and cost savings for a wide range of planetary missions. The present study analyzes the performance benefit offered by aerocapture for a set of design reference missions and their applications to future Solar System exploration from Venus to Neptune. The estimated performance benefit for aerocapture in terms of delivered mass increase are: Venus (92%), Earth (108%), Mars (17%), and Titan (614%), Uranus (35%), and Neptune (43%). At Uranus and Neptune, aerocapture is a mission enabling technology for orbit insertion from fast arrival interplanetary trajectories.

---

****** To whom correspondence should be addressed, E-mail: athulpg007@gmail.com

## I. INTRODUCTION

Aerocapture is a maneuver which uses aerodynamic drag to slow down a spacecraft in a single pass through the atmosphere to achieve nearly fuel-free orbit insertion [1, 2]. To date, all planetary orbiters have used propulsive orbit insertion. However, aerocapture is a promising alternative, especially for small satellite missions and missions to the ice giants [3]. The large $\Delta V$ requirement makes it practically impossible for small satellites to enter low circular orbits. Aerocapture can enable insertion of low-cost satellites into circular orbits around Mars and Venus [4]. For ice giant missions, aerocapture can enable orbit insertion from fast arrival interplanetary trajectories which are impractical with propulsive insertion due to the prohibitively large $\Delta V$ [5]. By utilizing the atmospheric drag to impart the $\Delta V$, aerocapture can offer significant propellant mass and cost savings for a wide range of missions [6]. The concept of operations for the aerocapture maneuver is shown in Figure 1, for a drag modulation vehicle at Mars. The aero-thermal conditions encountered during the maneuver depend on the destination, and performance benefit is also destination dependent [7, 8]. A recent NASA study underscored the need for design reference missions, as benchmarks for evaluating the benefits of aerocapture at various destinations. The present study uses the Aerocapture Mission Analysis Tool (AMAT) to analyze the performance benefit offered by aerocapture for a set of design reference missions [9, 10].

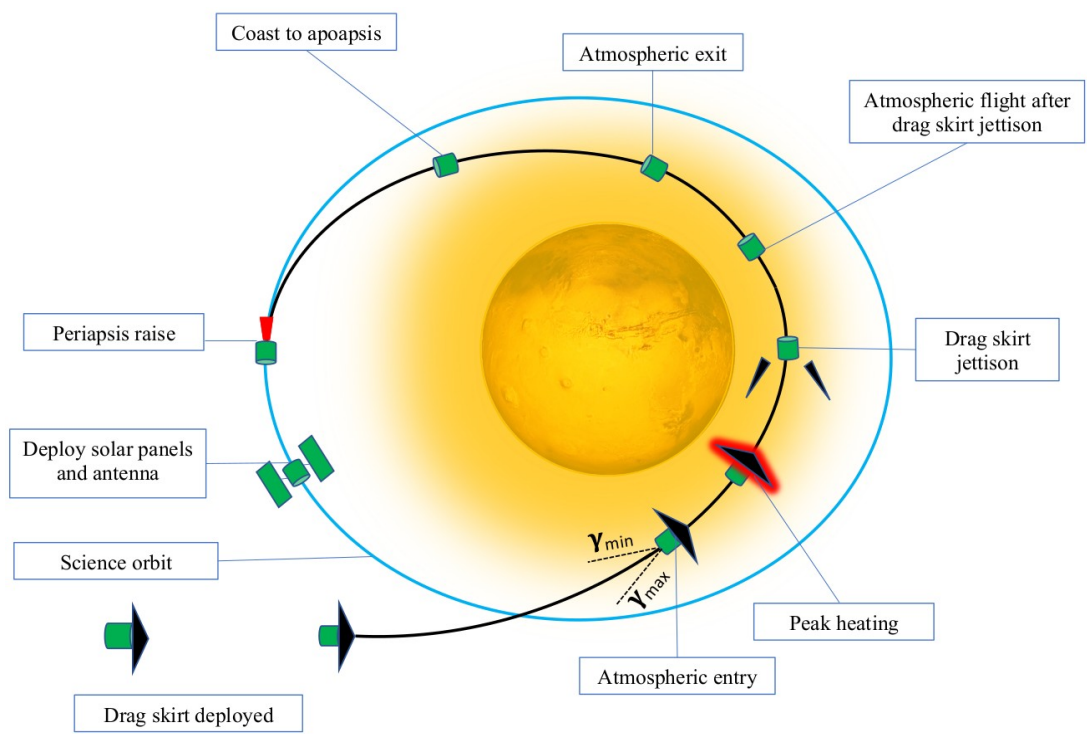

Figure 1. Aerocapture maneuver concept of operations.

## II. VENUS

Venus is Earth's closest planetary neighbor and has a thick atmosphere. The dense thick atmosphere also presents challenging entry conditions, making it not compelling for high-ballistic coefficient rigid entry vehicles [11, 12]. However, low-ballistic coefficient deployable entry systems such as ADEPT can greatly alleviate these difficulties and present an attractive method for inserting small satellites around Venus, particularly into low circular orbits for which the $\Delta V$ requirements are substantial. The deployable systems decelerate much higher up in the atmosphere, thus greatly reduces the aero-thermal heating. Recent work has established a design reference mission for inserting small satellites at Venus using drag modulation aerocapture [13]. The reference interplanetary trajectory arrives with a hyperbolic excess speed of 3.5 km/s and aims to insert a small 25 kg spacecraft into a 400 km circular orbit at Venus. The $\Delta V$ required for the propulsive orbit insertion maneuver is 3533 m/s, which is quite challenging to achieve for a small spacecraft propulsion system. Figure 2 compares the mass fraction delivered to orbit using propulsive insertion and aerocapture. With propulsive insertion, only 25% of the arrival mass can be inserted into orbit with the remaining 75% being the propulsion system. With aerocapture, about 50% of the arrival mass can be inserted into orbit. This has two important implications for small low-cost spacecraft. Aerocapture allows a 100% increase in delivered mass to orbit compared to propulsive insertion. Conversely, it can allow for smaller and cheaper spacecraft. A 25 kg orbiter at Venus would require launching a 100 kg wet mass spacecraft. Aerocapture would only require launching a spacecraft that is 50 kg at launch, smaller by a factor of 2. Preliminary estimates have indicated that by reducing the required $\Delta V$ from 3500 m/s (propulsive) to about 30 m/s (aerocapture, with periapsis raise maneuver), the mission cost can be reduced from over $100M to about $20M, a factor of 5, thus enabling a range of low-cost missions to Venus [14]. Examples include small standalone secondary ride-share payloads to Venus orbit, small satellites as part of large New Frontiers or Flagship missions [15], and missions to return atmospheric samples from the Venusian cloud layers [16].

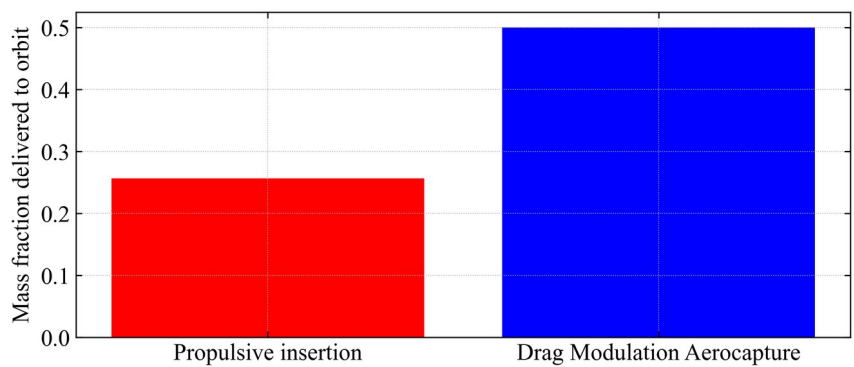

Figure 2. Performance comparison of propulsive insertion and aerocapture at Venus.

## III. EARTH

Aerocapture at Earth may be relevant for future sample return missions which seek to deliver samples to an orbiting space station rather than to Earth's surface for planetary protection reasons [17]. Aerocapture at Earth has also been studied for various technology demonstration experiments at Earth, although none were realized [18, 19]. Figure 3 shows the performance comparison of the delivered mass to a 400 km circular orbit from a trajectory with an excess speed of 3.5 km/s for which the orbit insertion $\Delta V$ is 3731 m/s. As with Venus, drag modulation aerocapture enables nearly a 100% increase in the mass delivered to orbit. Aerocapture at Earth also has applications to orbital transfer from GTO to LEO without the use of propellant, where it offers comparable performance advantages.

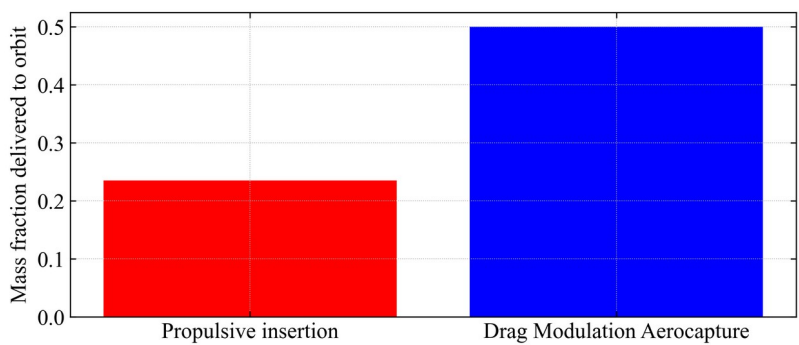

Figure 3. Performance comparison of propulsive insertion and aerocapture at Earth.

## IV. MARS

Figure 4 shows the performance comparison at Mars. The reference design seeks a 400 km circular orbit from an interplanetary trajectory with an excess speed of 2.6 km/s for which the orbit insertion $\Delta V$ is 2079 m/s. The performance benefit of aerocapture at Mars is considerably less that at Venus or Earth. However, the benign aero-thermal environment and make it an ideal candidate for a low-cost technology demonstration mission [20, 21].

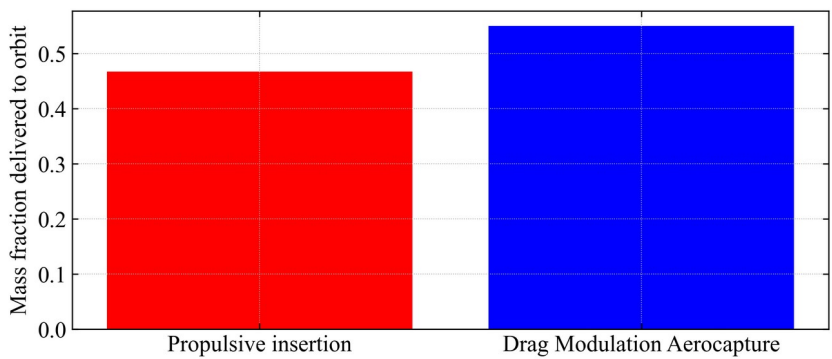

Figure 4. Performance comparison of propulsive insertion and aerocapture at Mars.

## V. TITAN

Titan's greatly extended thick atmosphere and the low entry speeds present an ideal destination for aerocapture. Titan offers the largest aerocapture corridor and the most benign aero-thermal environment of any Solar System destination. Ever since the Cassini-Huygens mission revealed Titan to have a diverse landscape and one of great scientific interest, there have been numerous mission proposals for a dedicated Flagship mission to Titan, with a Titan orbiter and a lander [22, 23, 24]. The Dragonfly mission will deliver a relocatable lander to Titan's surface. An orbiter around Titan remains to be accomplished by a future mission [25, 26]. However, getting into orbit around Titan requires very large ΔV, which requires enormous propellant and hence drives up the wet mass and mission cost. This has essentially precluded any New Frontiers or Discovery class mission concepts for a Titan orbiter, and with Europa and Uranus missions being the top priority for Flagships, a Flagship Titan mission is not viable in the near future. Drag modulation aerocapture offers an elegant solution to this challenge. The reference interplanetary trajectory arrives with a hyperbolic excess speed of 7.0 km/s and aims to insert a 2500 kg spacecraft into a 1700 km circular orbit at Venus. The ΔV required for the propulsive orbit insertion maneuver is 5832 m/s. This is so prohibitively large for a propulsive insertion, that only about 6% of the arrival mass can be delivered to orbit around Titan as shown in Figure 5. With aerocapture, 50% of the arrival mass can be delivered to orbit. This implies aerocapture offers enormous performance benefits for a future Titan orbiter mission, enabling a launch mass that is approximately 8 times smaller than what is possible with chemical propulsive insertion, potentially enabling a Titan orbiter to fit within the New Frontiers cost cap [27]. An spacecraft around Titan in a low-circular orbit can study Titan's surface with unprecedented detail using radar, mapping the entire surface at resolutions of 100s of meters and some regions at potentially much higher resolutions. Aerocapture is thus a key enabling technology for a future New Frontiers Titan orbiter mission.

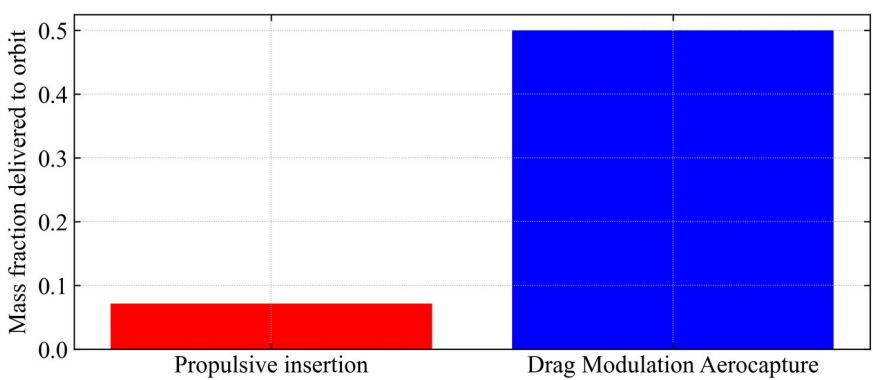

Figure 5. Performance comparison of propulsive insertion and aerocapture at Titan.

## VI. URANUS

The closer of the ice giants at 19 AU, Uranus is the top priority for a Flagship class mission in the next decade [28]. The large heliocentric distance of the ice giants poses significant mission design challenge to get there quickly, but also insert a reasonable payload into orbit. Due to risk considerations such as lack of confidence in the atmosphere models (potentially more perceived risk compared to actual risk), current baseline Uranus mission architectures have not used aerocapture [29, 30]. However, aerocapture has been shown to offer significant benefits for future Uranus missions [31, 32]. Two design reference missions are considered here. The first is a slow arrival (vinf = 10 km/s) trajectory, and the second is a fast arrival trajectory (20 km/s), both targeting a 4000 x 1M km orbit. For the slow arrival trajectory, the orbit insertion ΔV is 2667 m/s, and the ΔV for the fast arrival trajectory is 8631 m/s. With the slow arrival trajectory, drag modulation aerocapture is chosen as it is better suited compared to lift modulation due to corridor width and heating considerations. With the fast arrival trajectory, lift modulation aerocapture is chosen as it offers more corridor width and can use the HEEET TPS. Figure 6 shows the mass fraction delivered to orbit for the slow and fast arrival trajectories. For the slow arrival trajectory, drag modulation aerocapture is able to deliver about 35% more mass compared to propulsive insertion. For the fast arrival trajectory, the ΔV is so high that it is prohibitive for propulsive insertion. However, lift modulation aerocapture with an MSL-like aeroshell is still able to deliver 50% of the arrival mass to orbit. The fast arrival trajectory does present challenges associated with large heat loads in the range of 200–300 kJ/cm2, but HEEET is expected to be able to accommodate such large heat loads within a TPS mass fraction of about 25% [33, 34]. Figure 6 shows the enormous benefit offered by aerocapture for Uranus missions with fast arrival trajectories, enabling significantly shorter time of flight missions, with a reasonable payload mass fraction.

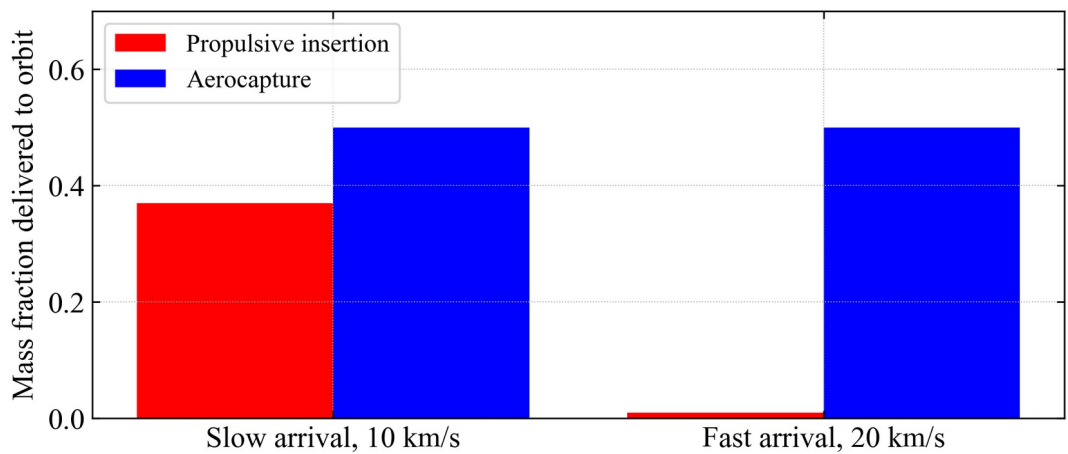

Figure 6. Performance comparison of propulsive insertion and aerocapture at Uranus.

## VII. NEPTUNE

The farther of the ice giants, Neptune is a more demanding destination for orbiter missions than Uranus [35]. Even though Uranus and Neptune are both scientifically compelling, the greater mission design challenges associated with Neptune appears to be the primary reason Uranus is preferred for the next Flagship mission. In contrast to Uranus, Neptune also offers the ability to study Triton, a captured Kuiper belt object which may be an active ocean world up close. Neptune aerocapture has been studied since 2003 using a mid-L/D vehicle to compensate for the large uncertainties [36]. However, since it has become clear such a vehicle would not be viable and recent studies have investigated using innovative techniques to leverage low-L/D aeroshells [37, 38, 39]. Two design reference missions are considered here. The first is a slow arrival (vinf = 10 km/s) trajectory, and the second is a fast arrival trajectory (20 km/s), both targeting a 4000 x 500,000 km orbit which is close to that of Triton. For the slow arrival trajectory, the orbit insertion ΔV is 2798 m/s, and the ΔV for the fast arrival trajectory is 8452 m/s. Figure 7 shows the mass fraction delivered to orbit for the slow and fast arrival trajectories at Neptune. For the slow arrival trajectory, drag modulation aerocapture is able to deliver about 40% more mass compared to propulsive insertion. For the fast arrival trajectory, the ΔV is again so high that it is prohibitive for propulsive insertion. However, lift modulation aerocapture with an MSL-derived aeroshell is still able to deliver 50% of the arrival mass to orbit. As with Uranus, Figure 7 shows the enormous advantage offered by aerocapture for Neptune missions with fast arrival trajectories. For ice giants, aerocapture essentially removes the upper limit on the arrival v_inf of about 12 km/s imposed by propulsive insertion. This opens up entirely new class shorter time of flight, fast arrival trajectories, making aerocapture an enabling technology for delivering well-instrumented orbiters to Uranus and Neptune with these fast trajectories [40].

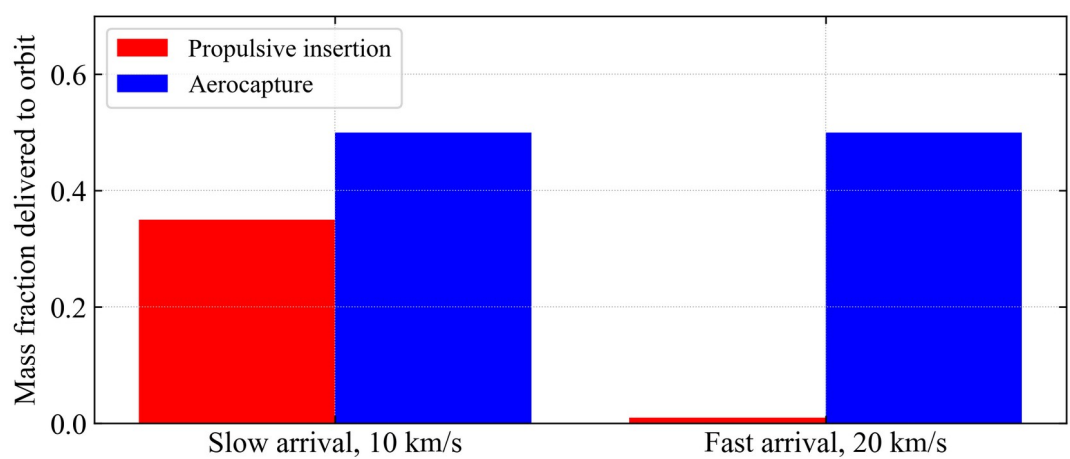

Figure 7. Performance comparison of propulsive insertion and aerocapture at Neptune.

## VIII. SUMMARY

Figure 8 summarizes the performance benefit of aerocapture across the Solar System destinations for the design reference missions. For Venus and Earth, drag modulation aerocapture provides nearly a 100% increase in delivered mass to a 400 km circular orbit compared to purely propulsive insertion. At Mars, the performance benefit is smaller at about 17%, but still significant. At Titan, aerocapture provides a 600% increase in delivered mass to a 1700 km circular orbit. At Uranus, for the slow arrival trajectories aerocapture provides a 35% increase in delivered mass to a 4000 x 1M km orbit compared to propulsive insertion. At Neptune, for the slow arrival trajectories aerocapture provides a 43% increase in delivered mass to a 4000 x 500,000 km orbit compared to propulsive insertion. At Titan, Uranus, and Neptune, aerocapture is a mission enabling technology for orbit insertion from fast arrival trajectories.

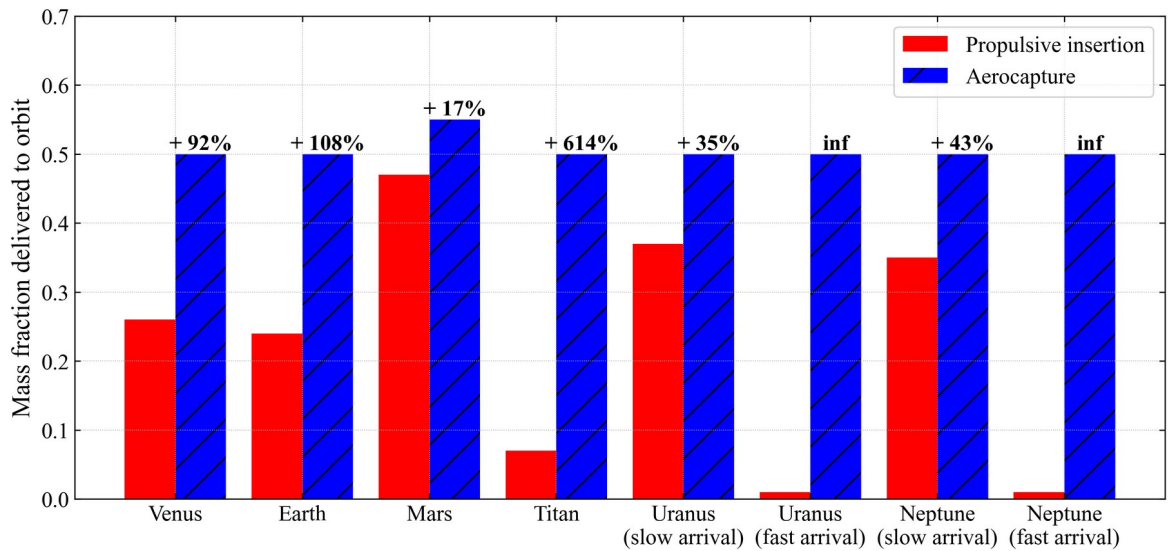

Figure 8. Summary of the aerocapture performance comparison.

## IX. CONCLUSIONS

The present study analyzed the performance benefit offered by aerocapture for a set of design reference missions. The estimated performance benefit of aerocapture in terms of delivered mass increase are as follows: Venus (92%), Earth (108%), Mars (17%), and Titan (614%), Uranus (35%), and Neptune (43%). At Titan, Uranus, and Neptune, aerocapture is a mission enabling technology for orbit insertion from fast arrival trajectories.

## DATA AVAILABILITY

The results presented in the paper can be reproduced using the open-source Aerocapture Mission Analysis Tool (AMAT) v2.2.22. The data and code used to make the study results will be made available by the author upon request.